\documentclass[one-column]{aastex631}

\newcommand\kepler{\textit{Kepler}}

\definecolor{sof}{rgb}{0.4, 0.0, 0.6}

\definecolor{cjg}{rgb}{0.0, 0.5, 0.5}

\usepackage[nolist]{acronym}
\usepackage{natbib}
\usepackage{wrapfig}

\submitjournal{The Astronomical Journal}
\accepted{November 24, 2022}

\newcommand{\PSUAA}{Department of Astronomy \& Astrophysics, 525 Davey Laboratory, The Pennsylvania State University, University Park, PA, 16802, USA}
\newcommand{\PSUCEHW}{Center for Exoplanets and Habitable Worlds, 525 Davey Laboratory, The Pennsylvania State University, University Park, PA, 16802, USA}
\newcommand{\PSETI}{Penn State Extraterrestrial Intelligence Center, 525 Davey Laboratory, The Pennsylvania State University, University Park, PA, 16802, USA}
\newcommand{\PSUIGC}{Institute for Gravitation and the Cosmos, The Pennsylvania State University, University Park, PA 16802, USA}

\newcommand{\BSRC}{Berkeley SETI Research Center, University of California, Berkeley, CA 94720, USA}

\newcommand{\UCBRAL}{Radio Astronomy Lab, University of California, Berkeley, CA 94720, USA}
\newcommand{\SETI}{SETI Institute, 339 Bernardo Avenue, Suite 200, Mountain View, CA 94043, USA}

\newcommand{\Carnegie}{Earth and Planets Laboratory, Carnegie Institution for Science, 5241 Broad Branch Road, NW, Washington, DC 20015, USA}

\newcommand{\UT}{Department of Astronomy, University of Texas at Austin, Austin, TX, USA}

\newcommand{\TCNJ}{Department of Physics, The College of New Jersey, 2000 Pennington Road, Ewing, NJ 08628, USA}

\newcommand{\ICRAR}{International Centre for Radio Astronomy Research, 1 Turner Ave, Bentley WA 6102, Australia}

\newcommand{\Manchester}{Department of Physics and Astronomy, University of Manchester, UK}

\newcommand{\Malta}{Institute of Space Sciences and Astronomy, University of Malta}

\newcommand{\BI}{Breakthrough Initiatives, NASA Research Park, Moffett Field, CA 94035, USA}

\newcommand{\UOG}{University of Greenwich, School of Computing and Mathematical Sciences, Park Row SE10 9LS London, UK}

\begin{document}
\raggedbottom

\title{A Green Bank Telescope search for narrowband technosignatures between 1.1 -- 1.9\,GHz during 12 Kepler planetary transits}

\author[0000-0001-7057-4999]{Sofia Z. Sheikh}
\affiliation{\SETI}
\affiliation{\BSRC}
\affiliation{\PSETI}

\author[0000-0001-8401-4300]{Shubham Kanodia}
\affiliation{\Carnegie}
\affiliation{\PSUAA}
\affiliation{\PSUCEHW}
\affiliation{\PSETI}

\author[0000-0003-0790-7492]{Emily Lubar}
\affiliation{\UT}

\collaboration{10}{Graduate SETI Course at Penn State}

\author[0000-0003-4381-5245]{William P. Bowman}
\affiliation{\PSUAA}
\affiliation{\PSUIGC}

\author[0000-0003-4835-0619]{Caleb I. Ca\~nas}
\affiliation{\PSUAA}
\affiliation{\PSUCEHW}

\author[0000-0002-1743-3684]{Christian Gilbertson}
\affiliation{\PSUAA}
\affiliation{\PSUCEHW}

\author[0000-0003-2372-1364]{Mariah G. MacDonald}
\affiliation{\PSUCEHW}
\affiliation{\TCNJ}

\author[0000-0001-6160-5888]{Jason Wright}
\affiliation{\PSUAA}
\affiliation{\PSUCEHW}
\affiliation{\PSETI}

\collaboration{10}{The Breakthrough Listen Initiative}

\author{David MacMahon}
\affiliation{\UCBRAL}
\affiliation{\BSRC}

\author[0000-0003-4823-129X]{Steve Croft}
\affiliation{\UCBRAL}
\affiliation{\BSRC}
\affiliation{\SETI}

\author[0000-0003-2783-1608]{Danny Price}
\affiliation{\BSRC}
\affiliation{\ICRAR}

\author[0000-0003-2828-7720]{Andrew Siemion}
\affiliation{\BSRC}
\affiliation{\SETI}
\affiliation{\Manchester}
\affiliation{\Malta}

\author{Jamie Drew}
\affiliation{\BI}

\author{S. Pete Worden}
\affiliation{\BI}

\nocollaboration{10}{}

\author[0000-0003-4941-2201]{Elizabeth Trenholm}
\affiliation{\UOG}

\correspondingauthor{Sofia Sheikh}
\email{ssheikh@seti.org}

\begin{abstract}

A growing avenue for determining the prevalence of life beyond Earth is to search for ``technosignatures'' from extraterrestrial intelligences/agents. Technosignatures require significant energy to be visible across interstellar space and thus intentional signals might be concentrated in frequency, in time, or in space, to be found in mutually obvious places. Therefore, it could be advantageous to search for technosignatures in parts of parameter space that are mutually-derivable to an observer on Earth and a distant transmitter. In this work, we used the L-band (1.1--1.9 GHz) receiver on the Robert C. Byrd Green Bank Telescope (GBT) to perform the first technosignature search pre-synchronized with exoplanet transits, covering 12 Kepler systems. We used the Breakthrough Listen turboSETI pipeline to flag narrowband hits ($\sim$3~Hz) using a maximum drift rate of $\pm$614.4~Hz/s and a signal-to-noise threshold of 5 --- the pipeline returned $\sim 3.4 \times 10^5$ apparently-localized features. Visual inspection by a team of citizen scientists ruled out 99.6\% of them. Further analysis found 2 signals-of-interest that warrant follow-up, but no technosignatures. If the signals-of-interest are not re-detected in future work, it will imply that the 12 targets in the search are not producing transit-aligned signals from 1.1 -- 1.9 GHz with transmitter powers $>$60 times that of the former Arecibo radar. This search debuts a range of innovative technosignature techniques: citizen science vetting of potential signals-of-interest, a sensitivity-aware search out to extremely high drift rates, a more flexible method of analyzing on-off cadences, and an extremely low signal-to-noise threshold.

\end{abstract}


\section{Introduction} 
\label{sec:intro}

A vast and still-growing part of our astronomical exploration is the search for life elsewhere in the universe. Many programs look for such life on exoplanets through their \textit{biosignatures}, surface features \citep[e.g.,][]{coelho2022color}, or atmospheric constituents \citep[e.g.,][]{thompson2022case} that indicate the presence of biological activity. Another complementary strategy is to look for the \textit{technosignatures} of technologically-capable life --- \ac{ETA}s, to use the terminology of \citet{Dobler2021} --- which may be more abundant, long-lived, highly-detectable, and unambiguous than other previously-described biosignatures \citep{wright2022case}.

The most popular technosignature search strategy to date is radio searches for artificial emission \citep[as pioneered by ][]{Drake1961}, which has grown exponentially in recent years, making use of cutting-edge computational techniques and newly-developed hardware \citep[e.g.,][]{harp2018application, ma2022first}. However, even with the renewed observational energy, the search space remains mostly unexplored \citep{Wright2018a}. This provides an opportunity for radio observation projects, large and small, to make an impact by filling in unexplored regions of parameter space.

One suggestion on how to best navigate this huge parameter space is to use ``Schelling Points'' \citep{schelling_strategy_1960, wright2020planck}, to prioritize mutually-derivable parts of parameter space which a transmitter and receiver can both land upon without any prior communication. This allows for more efficient traversal of parameter space --- potentially leading to a technosignature detection much sooner --- and also can be more power efficient for the transmitter, which can focus its energy in particular directions and times. One application of this idea is to prioritize certain places and times synchronized with an astronomical event \citep{pace1975time}. In some of these synchronized strategies, the event is external to the transmitter and receiver's systems, i.e., a nearby supernova or nova \citep{tang1976supernovae, lemarchand1994passive, 2022arXiv220604092D}. In another application, the synchronizing event is some feature of the transmitter or receiver's system, observable or predictable by both parties \citep[e.g.,][]{corbet2003synchronized}.

Here, we make use of exoplanetary transits as temporal Schelling points \citep{kipping_cloaking_2016}: if an 
\ac{ETA} transmitter on an exoplanet sends a signal towards its anti-stellar point, the signal will necessarily arrive to any observer along the line-of-sight at the same time as the exoplanet appears to be at mid-transit. This provides a known time and place (during an exoplanetary transit) for the observer to look for a signal. The transmitter may be targeting a specific system that happens to lie within its ecliptic, and thus sends a signal once each of its exoplanetary years. Conversely, the transmitter may be constantly targeting its exoplanet's anti-stellar point, sending out a transmission which sweeps across its ecliptic. In the extreme case, a tidally-locked planet hosting a single surface-locked transmitter could be constantly sweeping the ecliptic via the anti-stellar point, perhaps powered by energy collected photo-voltaically from the star-facing side of the exoplanet. Observations at this special periodic epoch offer a way to sample the large possibilities of repetition rates associated with periodic transmissions \citep{wright_how_2018}. Completely divorced from this Schelling Point logic, ETAs may preferentially emit high-power microwave radiation in their ecliptics (as we do in our solar system due to space communications), making transiting systems potentially more favourable for detecting \textit{unintentional} artificial signals.

In this work, we perform the first radio technosignature search of exoplanet-hosting stars, where observations were pre-synchronized with their transits. This is a complementary approach to that performed by \citet{franz2022breakthrough}, which looked back into archival data to identify serendipitous observations during transit. We also follow the growing tradition of conducting technosignature work and obtaining novel astronomical results using a cohort-based, academic course centered research model \citep{margot2021search, tusay2022seti}.

In Section \ref{sec:target_selection}, we describe how the targets were chosen. In Section \ref{sec:obs_and_data} we discuss the observing plan, the observation parameters, and the data format. We cover the methods for the search, including the software, assumptions, and chosen search parameters in Section \ref{sec:search_methods_data_reduction}. We discuss our findings and upper limits in Section \ref{sec:results}. Finally, we discuss and conclude in Section \ref{sec:discussion} and Section \ref{sec:conclusion} respectively.

\section{Target Selection}
\label{sec:target_selection}
\begin{figure}
    \centering
    \includegraphics[width=0.7\textwidth]{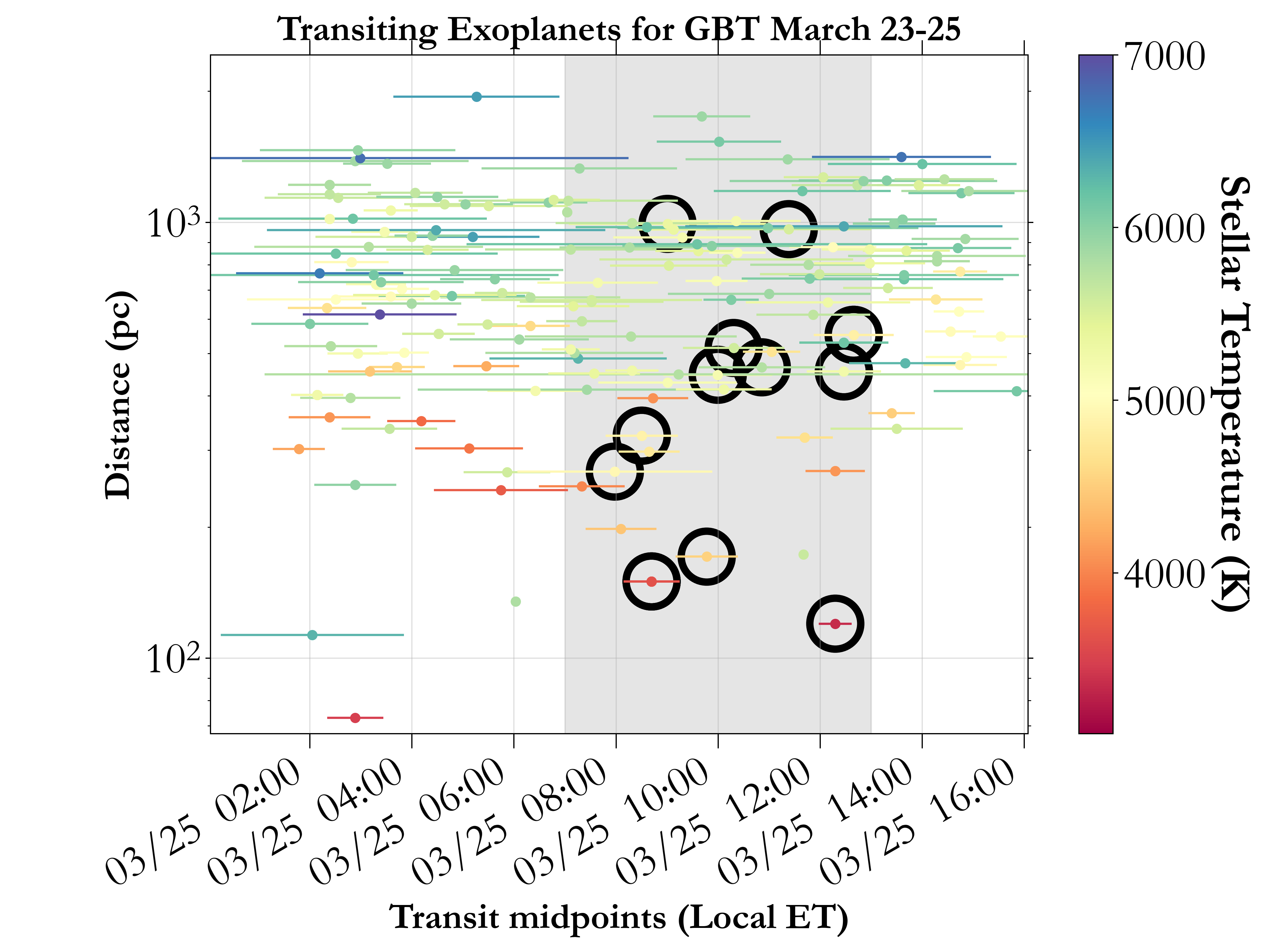}
    \caption{Confirmed transiting exoplanets in the Kepler field as observable from the Green Bank Telescope (GBT) during \replaced{a 2 day period in March 2018}{on March 25, 2018}. The horizontal errorbars depict the transit duration, whereas the colour of the markers represent the stellar effective temperature. Points without visible errorbars have extremely well-defined transit midpoints. \added{The grey region marks our observing window, whereas the targets observed are circled in black.} We did not include a habitability requirement in planet selection.}
    \label{fig:TransitVDistance}
\end{figure}

\begin{figure}[ht]
    \centering
    \includegraphics[width=0.6\textwidth]{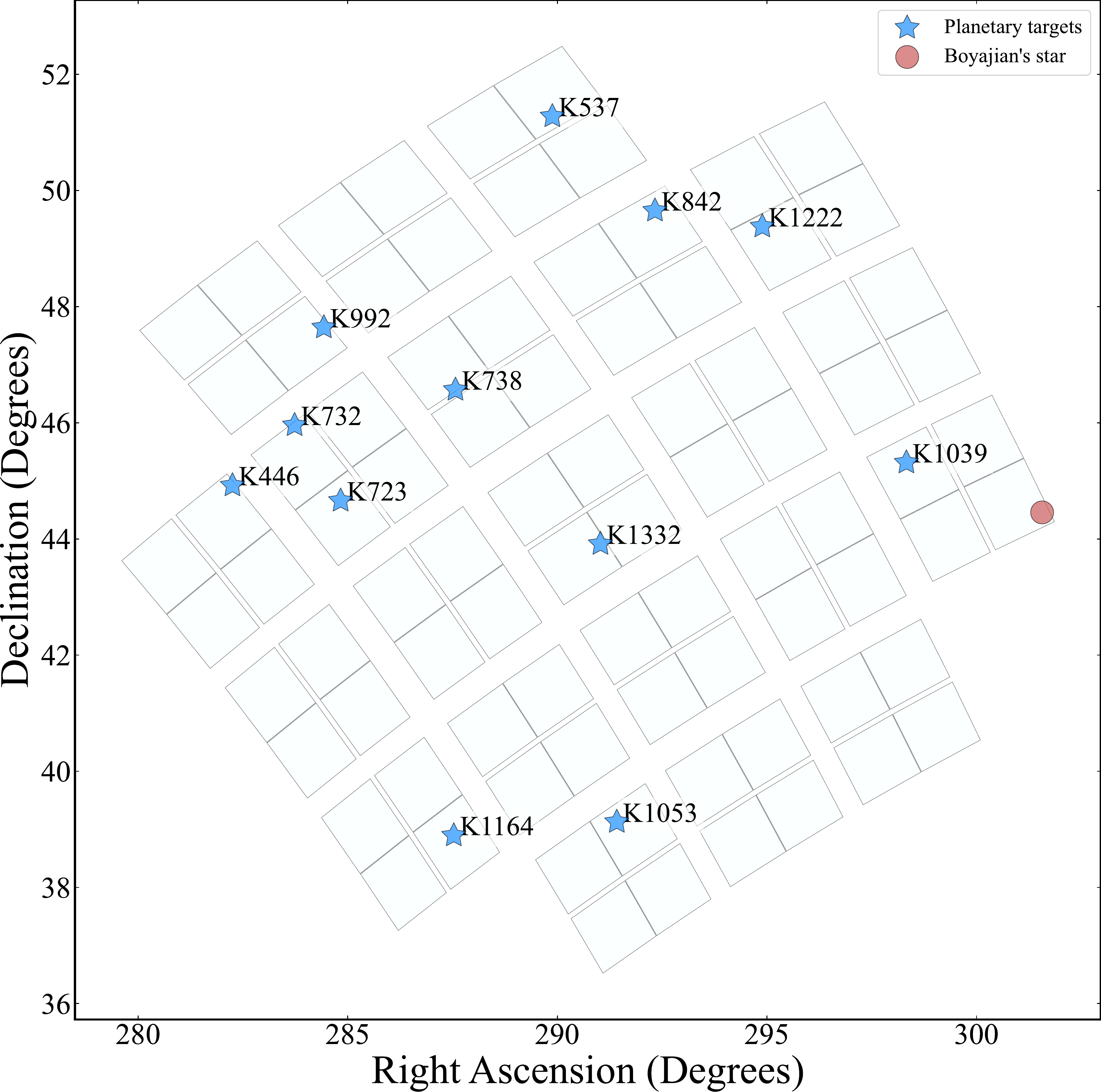}
    \caption{A plot of all exoplanetary systems (blue stars) observed with the GBT for this project, overlaid on the Kepler field of view, which covers 115 square degrees on sky \citep{Mullally2016}. The name of each system as it appears in Table~\ref{tab:observation_table} is shown. Boyajian's star (an off-source target) is also included for reference.}
    \label{fig:koifov}
\end{figure}

We first compiled a list of \replaced{all}{540} exoplanets that transit during a 2-day potential observing window \added{(March 23, 2018 -- March 25, 2018)} as a function of distance, using ephemerides from the NASA Exoplanet Archive \citep[Figure \ref{fig:TransitVDistance};][]{akeson_nasa_2013_doi}. Down-sampling from this subset of transiting exoplanets, we selected \added{436} planets discovered by the \textit{Kepler} mission due to their limited range of celestial coordinates. This enabled us to minimize our slew times during the nodding sequence, and boost the observing duty cycle. While there have been previous searches focused on transiting exoplanets from the \kepler{} mission \citep{harp_seti_2016, siemion_11-19_2013}, these have not prioritized observations during planetary transits. As explained above, we consider the transit a temporal Schelling point and hence aim to maximize the fraction of transit that we observed. To further improve our observing efficiency, we decided to observe alternate transiting planets as part of the on-off-on sequence used to identify \ac{RFI}, and identified pairs of exoplanets in the \kepler{} field that are transiting at similar times and are particularly close on-sky\footnote{None of these pairs are close enough on the sky to potentially cause source confusion, i.e., the angular distance between the targets $\gg$ GBT beam diameter in L band.}. \added{Of the \kepler{} transiting planets in our observing window we down-selected to 12 which are a) fairly close in their positions on-sky to allow for efficient alternating (on-off) exposures, and b) would result in sufficient sampling both in-and-out of transit, to fill the 6 hours of accepted observing time from \ac{BL}.}. The properties of our 12 observed targets are listed in \autoref{tab:planets}, and their on-sky positions are visualized in Figure \ref{fig:koifov}.

% finalized our list of transiting exoplanets during our scheduled observing block

\begin{table}[!b]
    \centering
    \small
    \caption{Stellar and planetary properties for the twelve transiting Kepler planets observed in this work. We use the properties from \citet{morton_false_2016} for all planets, except Kepler-446b which we pull from \citet{muirhead_kepler-445_2015}.}
    \label{tab:planets}
    \begin{tabular}{l c c c c c c c c}
    \hline
    \hline
    Target	&	RA	&	Dec	&	Distance 	&	Radius 	&	Period 	&	$t_0$ 	&	$T_{eq}$  &	$T_{eff}$	\\
    	&	(J2000)	&	(J2000)	& (pc)	&	 ($R_{\oplus}$)	&	 (d)	&	 (BJD - 2454000)	&	 (K) &	 (K)	\\
    \hline							
    Kepler-446b	&	18h49m0.05s	&	44d55m15.96s	&	120	&	1.5	$\pm$0.25	&	1.57	&	965.91	&	648	&	3464	\\
    Kepler-537b	&	19h19m31.20s	&	51d16m48.00s	&	465	&	1.41$^{+0.06}_{-0.04}$	&	3.25	&	1004.28	&	1181	&	5703	\\
    Kepler-723b	&	18h59m19.32s	&	44d39m29.30s	&	965	&	12.19$^{+1.58}_{-0.79}$	&	4.08	&	1002.64	&	1016	&	5655	\\
    Kepler-732c	&	18h54m55.68s	&	45d57m31.57s	&	150	&	1.27$^{+0.07}_{-0.1}$	&	0.89	&	967.54	&	893	&	3582	\\
    Kepler-738b	&	19h10m16.73s	&	46d34m4.30s	&	992	&	2.5$^{+0.2}_{-0.17}$	&	24.59	&	1006.70	&	530	&	5474	\\
    Kepler-842b	&	19h29m16.80s	&	49d38m60.00s	&	552	&	1.6$^{+0.08}_{-0.09}$	&	1.22	&	966.46	&	1252	&	4838	\\
    Kepler-992b	&	18h57m43.20s	&	47d38m24.00s	&	268	&	1.62$^{+0.04}_{-0.27}$	&	20.16	&	977.34	&	510	&	4975	\\
    Kepler-1039b	&	19h53m16.80s	&	45d18m36.00s	&	324	&	1.46$^{+0.09}_{-0.22}$	&	0.93	&	964.58	&	2080	&	4777	\\
    Kepler-1053b	&	19h25m40.80s	&	39d7m48.00s	&	171	&	0.98$^{+0.05}_{-0.04}$	&	2.41	&	965.43	&	953	&	4507	\\
    Kepler-1164b	&	19h10m7.20s	&	38d53m24.00s	&	447	&	1.12$^{+0.04}_{-0.07}$	&	3.98	&	966.62	&	936	&	5143	\\
    Kepler-1222b	&	19h39m33.60s	&	49d22m48.00s	&	455	&	0.79	$\pm$0.06	&	1.92	&	965.47	&	1246	&	5083	\\
    Kepler-1332b	&	19h24m7.20s	&	43d54m36.00s	&	465	&	1.37$^{+0.09}_{-0.05}$	&	11.87	&	973.21	&	728	&	5523	\\
    \hline
    \end{tabular}

\end{table}

\section{Observations and Data}
\label{sec:obs_and_data}

We pre-planned our observing sequence to ensure that we would hit each exoplanet during its transit, using individual 5-minute observing blocks spanning our 6 hour observing window on March 25, 2018 (starting at 11:00 UT); we elected to use the standard \ac{BL} \ac{GBT} integration of 5 minutes, and assumed a 2 minute overhead to account for e.g., the slewing and settling of the telescope. Table~\ref{tab:observation_table} shows the log of observations, including their relative temporal position to the target exoplanet's mid-transit point. We created individual GBT observing scripts for each pair which toggled back-and-forth between the two targets until they were replaced by the next target script. This sequence was adjusted dynamically throughout the observing session to select targets closest to mid-transit, in light of actual slew times and unanticipated observing overheads. For example, we started our transit observations with Kepler-992b for scans 0010 and 0012, the former of which spanned the transit midpoint. During these scans, we observed Kepler-738b as our `off' target. After covering the transit midpoint for Kepler-992b, we switched to Kepler-1039b and Kepler-732c, which had the next occurring transit midpoints. The total in-transit time for each planet in the sample varied in duration from 0.65--3.9 hours (median: 1.8 hours), so the 5-minute scans covered approximately 5\% of each transit.

We followed a similar logic through the rest of our observing window, where we tried to observe targets for at least 2 scans each during transit; furthermore, we prioritized observations of transit midpoints and tried to minimize slew times.  We bracketed our observing block with calibration observations of quasar 3C\,295 (scans 0006 and 0007) in the beginning, and pulsar B0329+54 (scan 0059) at the end. Additionally, we obtained one scan (scan 0009) of KIC\,8462852, commonly known as Boyajian's star \citep{boyajian_first_2018} before starting our transit sequence; it was a conveniently-located off-source also targeted by \ac{BL} laser technosignature searches using Lick Observatory's Automated Planet Finder telescope \citep{lipman_breakthrough_2019}. 

Data were recorded using the L-band receiver (1.1--1.9~GHz) \citep{gbtobservers2022} and the \ac{BL} backend on the \ac{GBT}, which in 2018\footnote{It is now capable of digitizing up to 12 GHz of instantaneous bandwidth.} was capable of digitizing up to 6 GHz of instantaneous bandwidth in 2 polarizations at 8-bit resolution --- for more information, see \citet{MacMahon2018}. The raw voltage data is then channelized using the \texttt{gpuspec} code to produce three spectral \texttt{SIGPROC} filterbank files containing Stokes I total power/frequency/time datacubes, each file at a different time and spectral resolution \citep{Lebofsky2019}. In the following analyses, we make use of the high spectral resolution data product, with a frequency resolution of $\sim 3$\,Hz and a time resolution of $\sim 18$\,s.

We performed data quality checks using the calibrator observations at the beginning and end of the observing session. The pulsar B0329+54 was easily visible in a \texttt{prepfold} plot \citep{ransom2011presto}, providing a first-order confirmation that the system was working as expected. In addition, it had an expected \ac{S/N} of 9172 given our system parameters, observing parameters, and its characteristics in the ATNF pulsar database\footnote{https://www.atnf.csiro.au/research/pulsar/psrcat/}, and was detected at an \ac{S/N} of 5306 (57.85\%). Given that pulsars in general (and this pulsar in particular) are not perfect flux calibrators (they exhibit variability in flux over time due to scintillation, which was distinctly present in the scan), this is entirely within the range of expected outcomes.

We also used the first 3C\,295 scan as a continuum flux calibrator ``on''-source, and the following observation of Boyajian's Star as an ``off'' source, to derive a system temperature. Assuming a flux density of 22.22\,Jy at 1408\,MHz \citep[as measured by ][]{ott1994updated}, and a spectral index of $\alpha=-0.7$, as found for the calibrator's hotspots in a recent LOFAR observation \citep{bonnassieux2022spectral}, we obtain an empirical system temperature measurement of $T_{sys} = 22.66$\,K. The theoretical value given in the GBT Proposer's Guide for the L-band receiver is $T_{sys} \approx 20$\,K, which is consistent with the experimental results.

\added{The data from this campaign is hosted at UC Berkeley, managed by the \ac{BL} project, and globally accessible via http at \url{https://bldata.berkeley.edu/kepler-transits/}. The repository includes the high-resolution spectral data product used for the following analysis, as well as the spectral line-resolution, and high-resolution data products from the same observations. Specifications for these other data products can be found in \citet{Lebofsky2019}.}

\section{Search Methods and Data Reduction}
\label{sec:search_methods_data_reduction}

\ac{BL} has a well-established data pipeline for performing narrowband \ac{SETI} searches on high frequency resolution filterbank files. This pipeline consists of import, plotting, and other utility functions in the \texttt{blimpy} package \citep{breakthrough2019blimpy,Price2019b}, and a narrowband search code \texttt{turboSETI} \citep{enriquez2019turboseti} based on the de-dispersion algorithm of \citet{Taylor1974} and adapted for de-Doppler searches. These pipelines have frequently been used for \ac{SETI} searches in the recent literature \citep[e.g., ][]{Siemion2013, smith2021radio}.

For this work, we ran \texttt{turboSETI}'s narrowband hit-finder with a maximum drift rate of $\pm$ 614.4\,Hz/s and a \ac{S/N} threshold of 5. Both of these parameters are unusual choices, for the reasons presented below. 

A typical drift rate for radio \ac{SETI} searches is $\sim\pm$10\,Hz/s, within a factor of a few \citep[e.g.,][]{sheikh2020breakthrough, franz2022breakthrough}. However, \citet{sheikh2019choosing} showed that known exoplanetary systems could produce drift rates up to $\pm$ 200\,nHz (i.e., $\pm200 \times \nu$\,Hz/s, where $\nu$ is the observing frequency in GHz). The largest drift rates would be expected from exoplanets with the largest radial accelerations relative to the receiver on Earth, e.g., transiting exoplanets that are viewed edge-on. This work marks the first time that a \ac{SETI} survey has followed the 200\,nHz recommendation, by using a maximum drift rate of $\pm$614.4\,Hz/s; this drift rate is sufficient to capture drift rates of $\pm$200\,nHz even at 1.9\,GHz, the highest frequency in these observations.

It should be noted that \texttt{turboSETI}, in its current configuration, does not \textit{automatically} account for the sensitivity loss sustained when incoherently-dechirping past the ``one-to-one point'' $\nu_{\mathrm{1-to-1}}$, where a signal drifts one frequency bin in every time bin \citep{sheikh2019choosing}. Here, we implement the first of the two partial solutions to this problem described by \citet{margot2021search}. We search the original high spectral-resolution filterbank file with a drift rate range from 0 to $\pm\nu_{\mathrm{1-to-1}}$\,Hz/s, in steps of $\sim 0.01$\,Hz/s. Then, we ``scrunch'' the file in frequency by a factor of 2, halving the number of bins by summing every other bin with its following neighbor. We then search again, using a new drift rate range of $\pm\nu_{\mathrm{1-to-1}}$ to $\pm2 \times \nu_{\mathrm{1-to-1}}$\,Hz/s (the new one-to-one point in the new file), with a correspondingly doubled drift rate step. We repeat this process until we have covered the entire desired drift rate range, which requires a series of 12 scrunch-and-search iterations. It should be noted that each frequency-scrunch, though it maximizes the \ac{S/N} within its range of drift rates, still causes an unavoidable $\sqrt{2}$ loss in sensitivity due to the original dimensions of the incoherent sum that produced the data product. The sensitivity losses are discussed further in Section \ref{ssec:upper_limits}, and it should be noted that exact signal positions and drift rates within the band may also cause irregularity in sensitivity with this method.

Recent \ac{SETI} searches have often chosen an \ac{S/N} of 10 for their hit thresholds, including \citet{price2020breakthrough} and \citep{margot2021search}. In narrowband radio signal detection these limits are primarily dictated by the filtering resources of the search, rather than any inherent statistical significance --- the environment is so contaminated by \ac{RFI} that false alarm rates from other common astronomical distributions, e.g., white noise or Gaussian statistics, do not apply. In this work, we instead decided upon a lower \ac{S/N} of 5, which allows us to double the sensitivity of the search. It should be noted that this causes an immense number of hits to pass the \texttt{turboSETI} filtering step, in addition to the shorter BAB cadence described in the following section: we managed this step with citizen science (Section \ref{ssec:citizen_scientists}), but also note its difficulties (Section \ref{sec:discussion}).

\section{Results}
\label{sec:results}

\begin{figure}
    \centering
    \includegraphics[width=\textwidth]{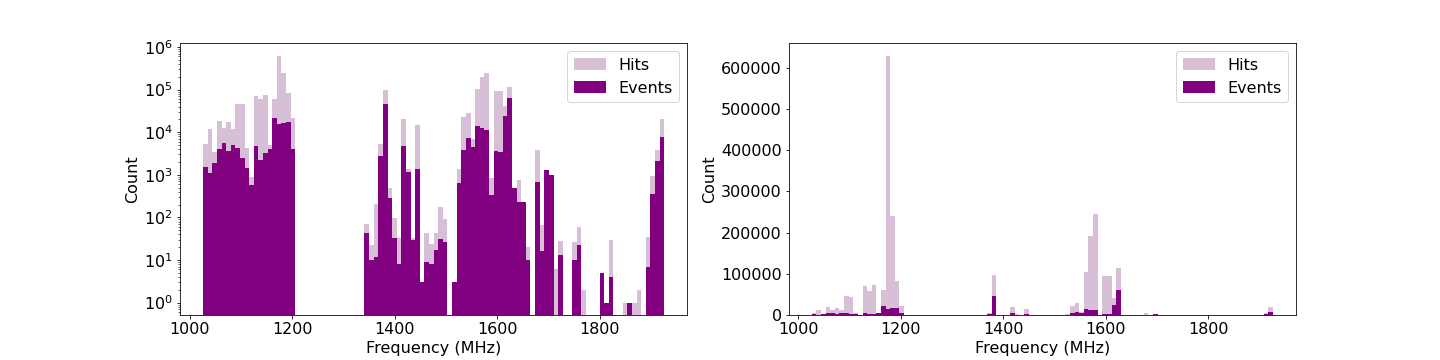}
    \caption{A histogram showing the frequency distribution of the hits (lighter) and events (darker) in a logarithmic (left) and a linear (right) scale. The notch filter is responsible for the absence of hits between 1200--1340~MHz. Three distinct frequency regions contain more than 100000 hits each: 1165--1185\,MHz, 1375--1385\,MHz, and 1570--1580\,MHz. These three regions are discussed in further detail in Section \ref{ssec:rfi_clusters}.}
    \label{fig:freq_hist_no_soi}
\end{figure}

We ran \texttt{turboSETI}'s hit-finding algorithm on every observation in Table \ref{tab:observation_table}. We ignored the region from 1.20--1.34~GHz, which corresponds to the L-band ``notch filter'', which applies to the most \ac{RFI}--contaminated region of the spectrum at the \ac{GBT}. This generated a total of $\sim$2.53 million hits. We then used \texttt{turboSETI}'s event-finding capability to compare hits in each Kepler observation (and Boyajian's star) to hits in each of the observations directly preceding and following it; this is the equivalent of an off-on-off or BAB cadence. In rare cases where the same target was observed twice in a row to realign the observations with the transit timeline (e.g., scans 0018 and 0019, both targeting Kepler-738b), we used the next closest scan in the preceding/following direction that was on a different target. This event-finding process resulted in 338473 unique events. The frequency distribution of the hits and events are shown in Figure \ref{fig:freq_hist_no_soi}.

\subsection{RFI-Heavy Bands}
\label{ssec:rfi_clusters}

We find that three frequency ranges --- 1165--1185\,MHz, 1375--1385\,MHz, and 1570--1580\,MHz --- contain the majority of the \ac{RFI} in our observations: 5\% of the band (excluding notch filter) accounts for 56\% of the hits. These ranges are consistent with the interference-heavy regions discussed by \citet{price2020breakthrough}. Here, we briefly discuss each of these frequency ranges in turn, and attempt to associate them with \ac{FCC} frequency allocations\footnote{\url{https://transition.fcc.gov/oet/spectrum/table/fcctable.pdf}}. %tried horizontal subfigs for this and they were too small

\begin{figure}
    \centering
    \includegraphics[width=\textwidth]{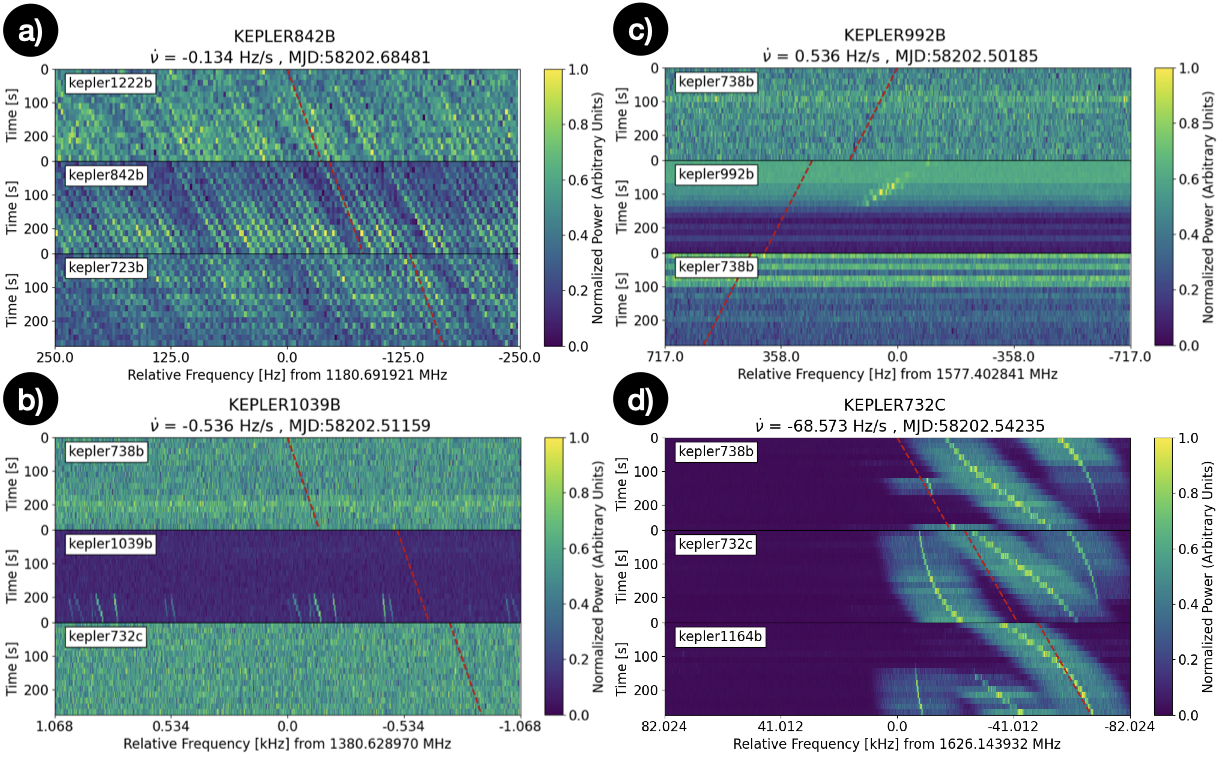}
    \caption{a) An example of an event in the 1165--1185~MHz region, consistent with \ac{GPS} L5. This event is \ac{RFI}, as it appears in multiple different targets in consecutive observations. The dashed red line represents \texttt{turboSETI}'s best-fit drift rate for the detected hit. b) An example of an event in the 1375--1385\,MHz region. These events occur for only a fraction of the observation (here, about 100 seconds) within a single scan --- this makes it difficult to determine whether they are impulsive \ac{RFI} or whether they are true transient signals localized on the sky. However, because we observe the same morphology of signal in multiple targets, and because of the match to the \ac{GPS} L3 downlink, we can assign these events as \ac{RFI}. c) An example of an event in the 1570--1580\,MHz region. These events are similar to the example shown in subfigure b in that they are degenerate between transient and localized signals; this is a common challenge for single-dish technosignature searches. The same signals were identified in multiple targets, however, indicating that they are indeed \ac{RFI} --- likely the \ac{GPS} L1 downlink. \added{d) An example of a high-drift rate signal around 1626\,MHz. We attribute this event to an Iridium satellite. It is clearly present in all three panels, and therefore \ac{RFI}, but the changing slope of the signals over time confounds the linear search algorithm.}}
    \label{fig:rfi}
\end{figure}

\textbf{1165--1185\,MHz}: There is an \ac{FCC} frequency allocation for aeronautical radionavigation and radionavigation-satellite (including space-to-Earth) transmissions between 1164--1215 MHz, covering this observed interference region in Figure \ref{fig:freq_hist_no_soi}. The distinct peak from 1165--1185\,MHz is consistent with the \ac{GPS} L5 downlink\footnote{\url{https://www.nist.gov/pml/time-and-frequency-division/popular-links/time-frequency-z/time-and-frequency-z-g}}. An example of an event in this region that passed \texttt{turboSETI}'s thresholds is shown in Figure \ref{fig:rfi}a.

\textbf{1375--1385 MHz}: This narrow region of the spectrum is occupied by the \ac{GPS} L3 downlink band, which provides communications links for the U.S. Nuclear Detonation Detection System. An example of an event in this region that passed \texttt{turboSETI}'s thresholds is shown in Figure \ref{fig:rfi}b.

\textbf{1570--1580\,MHz:} Once again, this interferer falls within an \ac{FCC} allocation dedicated to aeronautical radionavigation and radionavigation-satellite (including space-to-Earth) transmissions; in this case, however, the allocation is much wider (1559 MHz--1610\,MHz) than the region where we see the majority of interference. An example of an event in this region that passed \texttt{turboSETI}'s thresholds is shown in Figure \ref{fig:rfi}c. Given the frequency range, and the presence of other \ac{GPS} downlinks within the dataset, these hits can likely be attributed to the \ac{GPS} L1 downlink centered at 1575\,MHz.

Finally, we detected a series of swooping signals (changing from high drift rate, to low drift rate, to high again) between 1616--1626.5\,MHz that also passed \texttt{turboSETI}'s event filtering. \added{We highlight these signals due to their high drift rates, which reached an absolute value of nearly 70 Hz/s.}These signals account for the event spike above 1600\,MHz in the right panel of Figure \ref{fig:freq_hist_no_soi}. An example is shown in Figure \ref{fig:rfi}d. We attribute these signals to the Iridium satellite constellation's L-band downlink\footnote{\url{https://apollosat.com/iridium-satellite-frequency-bands}}, and note that these signals \added{are bright enough to} have passed \texttt{turboSETI} event filters in multiple \ac{GBT} L-band \ac{SETI} observations\added{, even those with lower drift rate bounds.} \citep[e.g., ][]{enriquez2017breakthrough, tusay2022seti}.

\subsection{Filtering By Citizen Scientists}
\label{ssec:citizen_scientists}

At this stage, with $>340,000$ events, most radio technosignature campaigns would make a change to the event thresholds to get them to a number suitable for visual inspection by a single researcher. For example, the \ac{S/N} could be raised from 5 to 10 (leaving only $92,000$ events), or even 25 (leaving only $26,000$ events), or the drift rate could be further reduced. However, making these changes would lower our sensitivity. For this campaign, we decided instead to apply the power of crowdsourcing via a limited citizen science project.

We created .pdf files containing 1000 plots each, making them accessible via cloud services such as Box and Google Drive, and developed an hour-long lecture and interactive quiz to train volunteers. Six undergraduate volunteers from the \ac{PSPSC} club\footnote{An affiliate of the Pulsar Search Collaboratory project described by \citet{blumer2020pulsar} and others.} looked through output plots such as those shown in Figure \ref{fig:rfi}, and flagged any signal that appeared in only the on-source middle panel. Due to the COVID--19 pandemic, the \ac{PSPSC} members only completed approximately 20\% of the sample in the Fall 2020 semester --- additional volunteers were recruited from the lead author's professional network to assist with signal filtering.

In the end, 13 citizen scientists (named in the Acknowledgements) flagged approximately 0.4\% ($\sim 1600$ signals) of the dataset for further analysis, reducing the number of interesting events to a few thousand. It should be noted that this number is approximate: some signals were identified multiple times at different drift rates (see Section \ref{sec:discussion}), and clusters of plots with repeating signals (such as those in Figure \ref{fig:rfi}a) were grouped as a single phenomenon. Due to the size of the dataset compared to the number of volunteers, we did not have the resources to send the same data to multiple recipients, as is done in projects such as Zooniverse\footnote{\url{https://www.zooniverse.org/}}. To ensure quality in flagging, the majority of the work was done in collaborative ``hack sessions'' with multiple participants on the same call, so plots-of-interest or borderline cases could be discussed and viewed by multiple participants simultaneously. The citizen scientists' effort meant that the filtered dataset was now at a manageable size for visual inspection by the authors, all while maintaining the extremely sensitive \ac{S/N} 5 threshold.

\subsection{Filtering Events with Context}

\begin{figure}
    \centering
    \includegraphics[width=\textwidth]{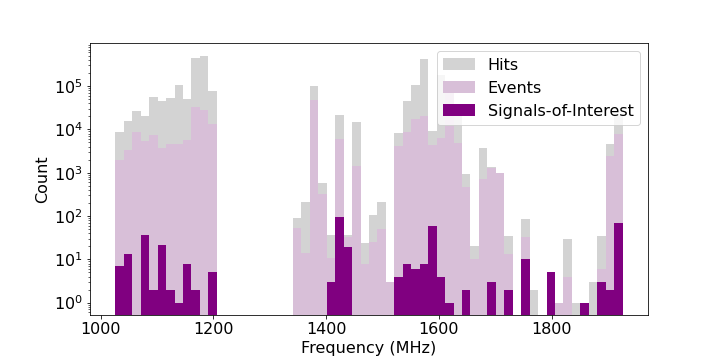}
    \caption{A histogram showing the frequency distribution of the hits (light gray), events (light purple), and signals-of-interest (purple). The signals-of-interest are distributed relatively evenly throughout the spectrum. Note the gap from 1.20--1.34\,GHz due to the GBT L-band receiver's notch filter.}
    \label{fig:sois_freq}
\end{figure}

\begin{figure}
    \centering
    \includegraphics[width=\textwidth]{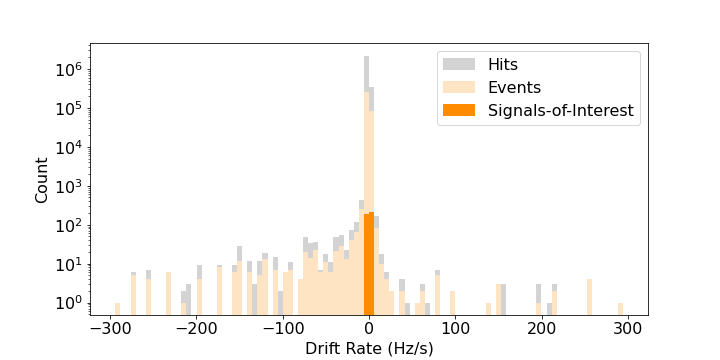}
    \caption{A histogram showing the drift rate distribution of the hits (light gray), events (light orange), and signals-of-interest (orange). The signals-of-interest are found only at low absolute drift rates.}
    \label{fig:sois_drift}
\end{figure}

\begin{figure}
    \centering
    \includegraphics[width=\textwidth]{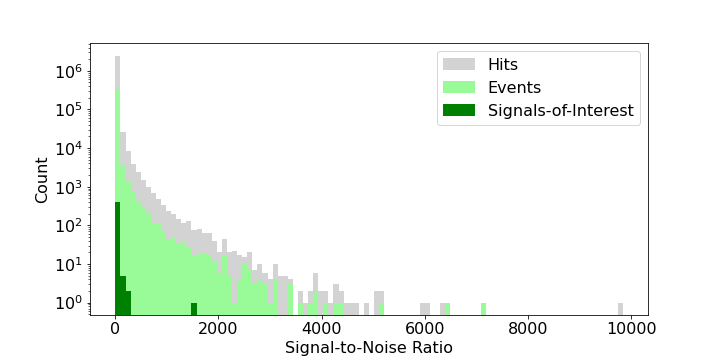}
    \caption{A histogram showing the \ac{S/N} distribution of the hits (light gray), events (light green), and signals-of-interest (green). The signals-of-interest are primarily faint. There are also 5 hits with S/N $>10000$ which are not shown on this plot for readability.}
    \label{fig:sois_snr}
\end{figure}

Each of these plots, in isolation, had the characteristics of a localized signal on the sky (only present in the on-source observation). However, following the technosignature verification framework of \citet{sheikh2021analysis}, these signals should also be analyzed in the context of similar, nearby signals. We grouped the remaining few thousand plots by frequency and morphology. Then, we checked to see if identical signals appear in two different targets (indicating that they are not local to either) or if we see a near-identical signal within a set that is definitely interference \citep[the way that blc1 was eventually disproved by][]{sheikh2021analysis}. With this strategy, we reduced the pool of signals-of-interest to 479. The frequencies, drift rates, and \ac{S/N}s of these 479 signals are shown in the context of the overall hits and events in Figures \ref{fig:sois_freq}, \ref{fig:sois_drift}, and \ref{fig:sois_snr} respectively.

It is, however, possible that the 479 signals-of-interest have disqualifying features \textit{elsewhere} in the observing session; for example, a sub-threshold repetition of the signal on a non-adjacent scan would not be picked up in the previous filtering step. Therefore, as a final verification, we plotted each of the remaining 479 plots in the context of the entire session of observations (all 52 observations in Table \ref{tab:observation_table}), to produce stacked waterfall plots with extreme aspect ratios. We eliminate a signal-of-interest from our list if it is a continuation or repetition of signals in other targets at any time during the morning of observations: this is a similar strategy to that applied in Section \ref{ssec:citizen_scientists} but now across the largest possible time baseline. 

\begin{figure}
    \centering
    \includegraphics[width=\textwidth]{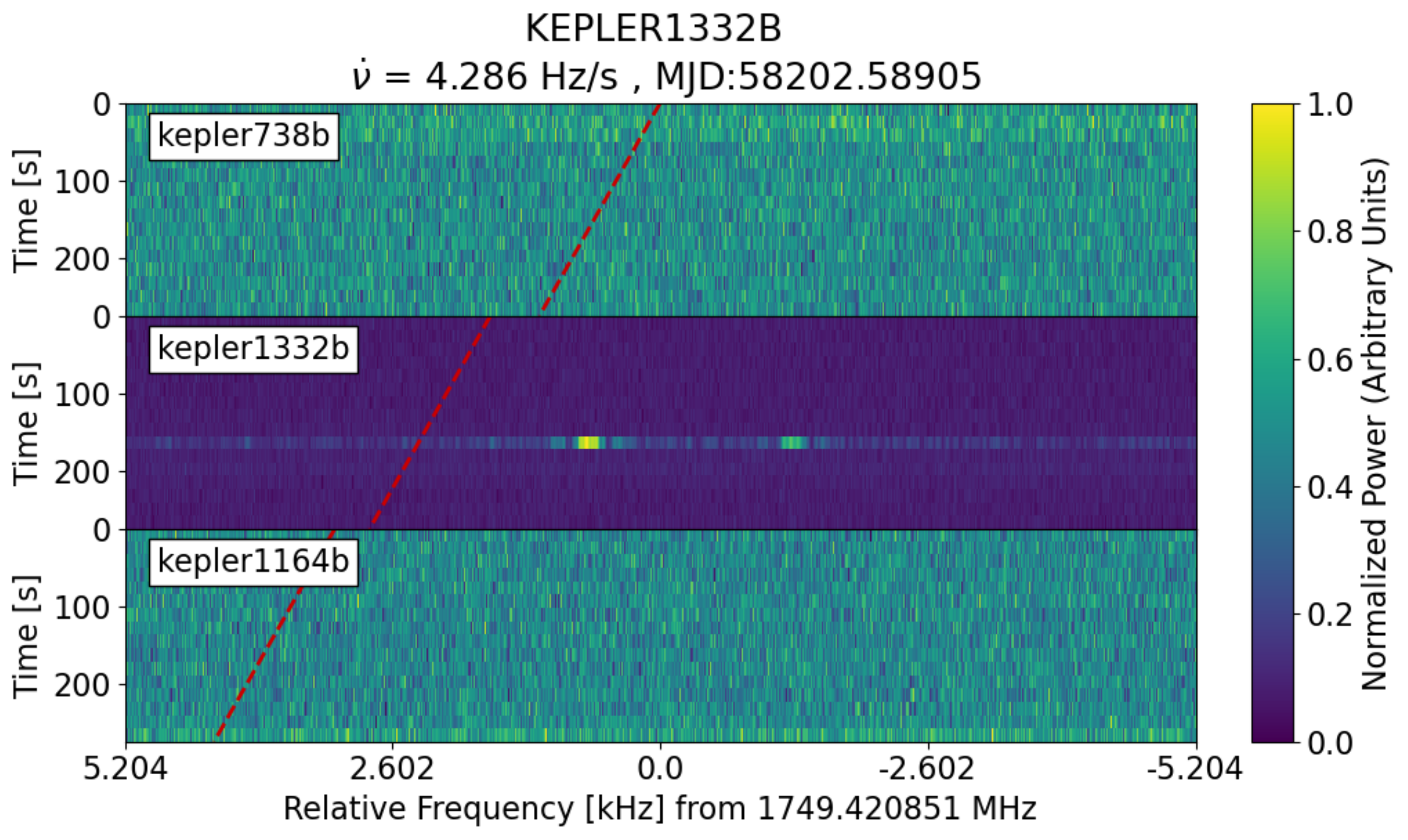}
    \caption{The signal-of-interest detected in Kepler-1332b.}
    \label{fig:1332_soi}
\end{figure}

\begin{figure}
    \centering
    \includegraphics[width=\textwidth]{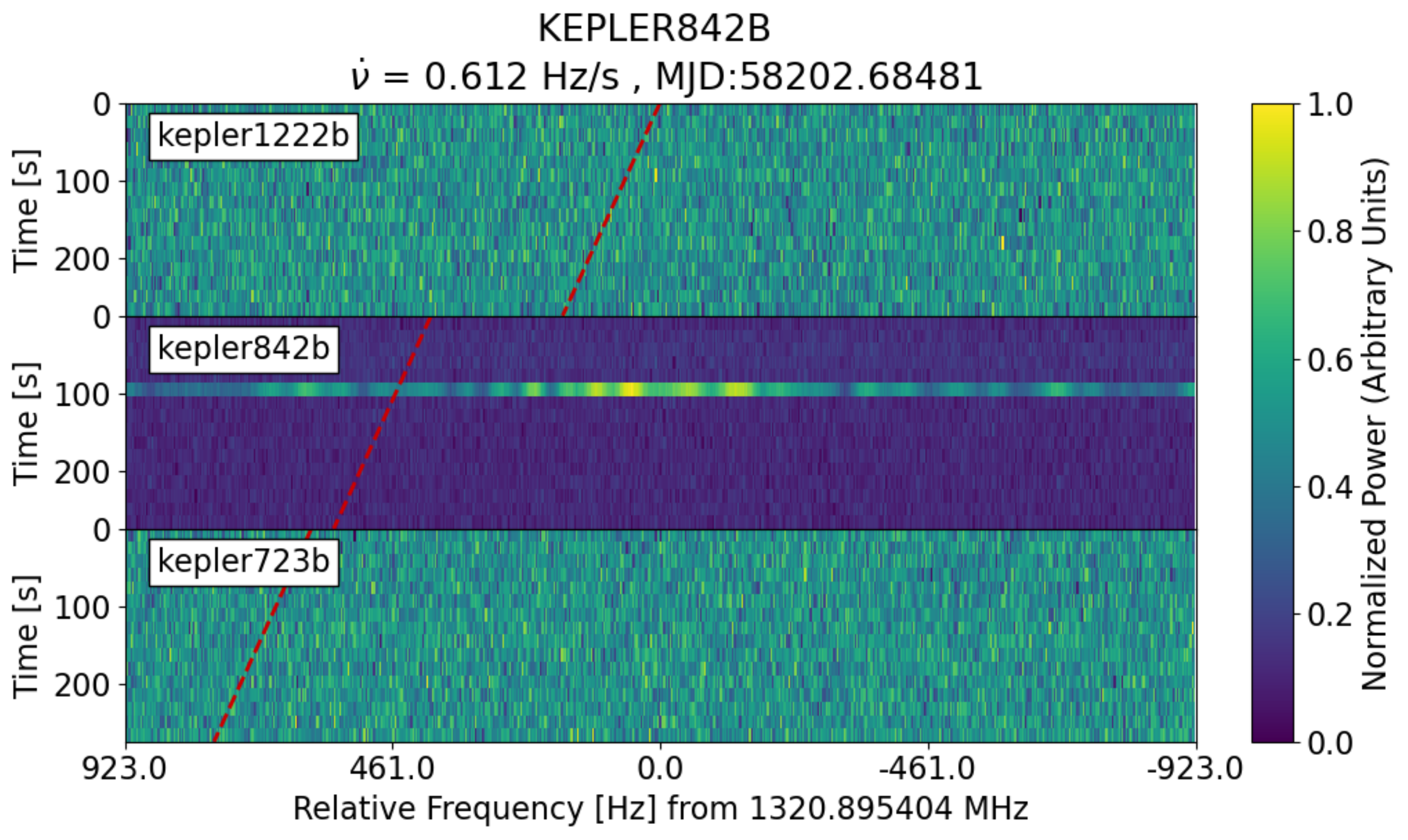}
    \caption{A portion of the signal-of-interest detected in Kepler-842b. This signal-of-interest spans 1040--1438 MHz, so this is not the full signal, but rather a narrow bandwidth example of one of the ``hits'' on this signal generated by \texttt{turboSETI}.}
    \label{fig:842_soi}
\end{figure}

This process eliminated all but 2 signals-of-interest: one in a scan of Kepler-1332b and one in a scan of Kepler-842b. Neither of these signals are actually narrowband, but they were bright enough to be detected by \texttt{turboSETI} and we did not want to restrict ourselves to only the signal morphologies we expected, if we serendipitously found a signal with a different morphology. Interestingly, both the detection in Kepler-1332b (during scan 0031) and the detection in Kepler-842b (during scan 0056) were during their respective transit midpoints. The waterfall plots are shown in Figures \ref{fig:1332_soi} and \ref{fig:842_soi}, and the signal properties are summarized in Table \ref{tab:signal_properties}.

\begin{table}[ht]
    \centering
    \begin{tabular}{|c|c|c|c|c|c|}
    \hline
    Target & Scan & MJD of Scan & Frequency & Transit Phase & Signal Type \\
    \hline
    Kepler-1332b & 0031 & 58202.59350 & 1749.4209 MHz & Transit Midpoint & Pulse \\
    Kepler-842b  & 0056 & 58202.69226 & 1040--1438 MHz & Transit Midpoint & Broadband Pulse \\
    \hline
    \end{tabular}
    \caption{Properties of the two interesting signals-of-interest reported in this study. \added{``Transit Midpoint'' indicates that the midpoint of the planet's transit occurred during this scan.}}
    \label{tab:signal_properties}
\end{table}

\added{As both signals-of-interest are contained within a single time bin in the high frequency-resolution product, we then checked the mid-resolution and high time-resolution products for evidence of the same signal. The high time-resolution did not reveal anything of interest, likely due to its low frequency-resolution, but the mid-resolution product ($\sim$3 kHz frequency resolution, 1 second time resolution) does provide additional information. The signal-of-interest in Kepler-1332b splits into two still-unresolved bursts spaced about one second apart. The signal-of-interest in Kepler-842b is also revealed to have a more complex structure in time, also periodic in frequency, which repeats across the entire bandwidth. These mid-resolution plots are shown in the Appendix. This new context about the morphology of the signal-of-interest lends more evidence to the idea that the signal-of-interest is broadband \ac{RFI}.}

We are close to the limit of what we can deduce from this dataset for these two signals-of-interest. With the information we have, it is impossible to exclude these signals as being due to \ac{RFI}\added{, which should be treated as the most likely explanation.}. Therefore, we recommend a follow-up observation of these two systems in L-band, during transit, with a different instrument. We are planning to do this with the newly-refurbished Allen Telescope Array (ATA) \citep[e.g.,][]{farah2021bright}. These signals-of-interest do not merit the scrutiny of blc1 \citep{sheikh2021analysis} because they only appear in a single 18 second integration; while this may be expected for synchronized transmitters which only send a signal at mid-transit, it also makes the standards-of-evidence higher (no proof of signal localized on sky, no drift rate measurement, etc.), requiring reobservation as the next step before more detailed analyses. 

\subsection{Upper Limits}
\label{ssec:upper_limits}

We calculate technosignature upper limits following the method of \citet{price2020breakthrough}. The minimum detectable flux density, $F_{\mathrm{min}}$ for a narrowband signal is given by:

\begin{equation}
    F_{\mathrm{min}} = \textrm{S/N}_{\mathrm{min}} \frac{2k_BT_{\mathrm{sys}}}{A_{\mathrm{eff}}} \sqrt{\frac{\delta \nu}{n_{\mathrm{pol}}t_{\mathrm{obs}}}}
\end{equation}

Here, $\textrm{S/N}_{\mathrm{min}}$ is the signal-to-noise threshold, $k_{B}$ is the Boltzmann constant, $T_{\mathrm{sys}}$ is the system
temperature, $A_{\mathrm{eff}}$ is the effective collecting area of the
telescope, $\delta \nu$ is the frequency channel width, and $n_{\mathrm{pol}}$ is the number of polarizations. We make the same assumption as \citet{price2020breakthrough}: that the transmitting signal is the same width as the frequency channel $\delta \nu$, so no additional constant factor is needed to downweight $F_{\mathrm{min}}$.

With an $\textrm{S/N}_{\mathrm{min}}$ of 5, a $T_{\mathrm{sys}}$ of 20\,K, an $A_{\mathrm{eff}}$ of 5655\,m$^2$ (equivalent to using an effective diameter of 100m for the \ac{GBT}, with an L-band aperture efficiency of 0.72), a $\delta\nu$ of 3 Hz, an $n_{\mathrm{pol}}$ of 2 and a $t_{\mathrm{obs}}$ of 300 seconds, we calculate an $F_{\mathrm{min}}$ of 3.45 $\times$ 10$^{-26}$~W/m$^2$, equivalent to 1.15~Jy. This is consistent with other \ac{BL} searches at L-band with the \ac{GBT} \citep[e.g., ][]{enriquez2017breakthrough, price2020breakthrough}, but with extra sensitivity due to the low signal-to-noise threshold.

The population of Kepler exoplanets is relatively distant, with all of the targets in this study falling between 150--992\,pc. The \ac{EIRP} is calculated as:

\begin{equation}
    EIRP_{\mathrm{min}} = 4 \pi d^2 F_{\mathrm{min}}
\end{equation}

where $d$ is the distance to the furthest target in the survey. Using 992\,pc, we get a survey-wide EIRP$_{\mathrm{min}}$ of 406\,TW, or $\sim 20 \times$ the EIRP of the Arecibo radar. The two targets that showed signals-of-interest, Kepler-1332b and Kepler-842b, are located closer, at 465\,pc and 552\,pc respectively. This leads to target-specific \ac{EIRP}s of 89\,TW and 125\,TW, or $4.5$--$6.3\times$ the EIRP of Arecibo.

Finally, one can also express an upper limit in the form of a fraction of the 8-dimensional ``haystack'', as described by \citet{Wright2018a}. Using the same minimum and maximum axis bounds as described in that work, we calculate a total haystack fraction of $4.42 \times 10^{-20}$ covered by this search, comparable to Project Phoenix searches with Parkes and Arecibo \citep[e.g.,][]{backus2002project}.

\section{Discussion}
\label{sec:discussion}

This search represents the lowest \ac{S/N} ever chosen for a Breakthrough Listen affiliated project on the \ac{GBT}. However, caution is advised if following our procedure. We found that lowering the \ac{S/N} to 5 made \texttt{turboSETI}'s temporal on-off filtering much less effective. The lower \ac{S/N} magnified the likelihood of e.g., a bright pixel in the ``on'' target causing a hit to be flagged as an event. It may be more effective to recoup sensitivity in other parts of the signal processing chain, rather than as a threshold in \texttt{turboSETI}. This may be especially relevant for high drift rate signals \citep[using methods such as boxcar convolution, e.g., ][]{adamek2020single}.

In addition, vetting $\sim 3.4 \times 10^5$ events involved a large amount of work, even considering the citizen science approach. Using a single on-source with two off-sources produces a degeneracy between intermittent \ac{RFI} and sky-localized signals in a single-dish telescope, compounding the issue. We recommend that future searches invest in a better understanding of their \ac{RFI} environment and try to algorithmically filter further before performing visual classification. 

One approach is to better characterize the frequencies, drift rates, and morphologies of common interference sources, and only down-select signals that are not particularly consistent with those properties. Another approach is to use employ spatial filtering with a multi-beam receiver (such as the \ac{GBT}'s K-band Focal Plane Array) or an interferometer (such as the \ac{ATA}).

\begin{table}[ht]
    \centering
    \begin{tabular}{|l|l|l|l|}
        \hline
        $\dot{\nu}_{\mathrm{max}}$ (Hz/s) & Sensitivity Loss Factor & Number of Hits in Bin [\% of total] & Number of Events in Bin [\% of total]\\
        \hline
        0.15   & $1$    &   1838435 [53.1\%]    &   228741 [54.7\%]\\
        0.3    & $1.4$  &   1216882 [35.1\%]    &   129808 [31.0\%]\\
        0.6    & $2$    &   376107 [10.9\%]     &   46390 [11.1\%]\\
        1.2    & $2.8$  &   21999 [0.64\%]      &   7569 [1.81\%]\\
        2.4    & $5.7$  &   5549 [0.16\%]       &   3232 [0.77\%]\\
        4.8    & $8$    &   2607 [0.08\%]       &   1514 [0.36\%]\\
        9.6    & $11.3$ &   816 [0.02\%]        &   441 [0.11\%]\\
        19.2   & $16$   &   251 [$<0.01$\%]     &   134 [0.03\%]\\
        38.4   & $22.6$ &   139 [$<0.01$\%]     &   68 [0.02\%]\\
        76.8   & $32$   &   172 [$<0.01$\%]     &   90 [0.02\%]\\    
        153.6  & $45.3$ &   125 [$<0.01$\%]     &   71 [0.02\%]\\
        307.2  & $64$   &   83 [$<0.01$\%]      &   54 [0.01\%]\\
        614.4  & $90.5$ &   0 [$<0.01$\%]       &   0 [$<0.01$\%]\\
        \hline
    \end{tabular}
    \caption{Maximum drift rates (absolute value), the corresponding multiplier for reduction in sensitivity, and the number of hits affected by that reduction. The reason for the sensitivity reduction is described in Section \ref{sec:search_methods_data_reduction}. 99\% of hits and 97\% of events that we detected suffer a sensitivity loss of 2 or less. We can also consider the implied population of high-drift hits that we did \textit{not} detect, due to the sensitivity loss. However, the number of high-drift rates is so many orders-of-magnitude below the low-drift hits, that we do not expect the number of ``missed'' hits to meaningfully affect the total number. This implies that the sensitivity loss at high-drift rates should not greatly affect the conclusions of our survey. Note that this logic assumes that high-drift hits are not preferentially likely to be caused by an \ac{ETA}.}
    \label{tab:scrunch_sensitivities}
\end{table}

Also, as \citet{margot2021search} has mentioned, the \texttt{turboSETI} pipeline does not optimize for signals with high drift rates by default, such that signals suffer from drift-smearing across frequency bins, lowering the sensitivity. However, we can recover as much sensitivity as is physically possible (in an incoherent data product) by performing stepped frequency scrunching. The remaining sensitivity loss factors for this data, and how many signals were affected by those factors, are shown in Table \ref{tab:scrunch_sensitivities}. While we recommend that other \texttt{turboSETI} users also perform frequency scrunching if their searches cover applicably high drift rates, we note that each scrunched data product must be searched separately. This means that bright-enough hits will be detected redundantly multiple times, at multiple drift rates, as the hits from each scrunched data product are produced independently. This provides a strong motivation to integrate drift-rate-stepped scrunching into \texttt{turboSETI} in a more robust way.

Another way to address high-drift signals is to coherently correct the raw voltage data to an accelerating reference frame or fiducial drift rate (e.g., $\pm$100\,Hz/s) before the creation of the reduced and channelized data products \citep[as done in pulsar de-dispersion, e.g., ][]{sobey2022searching}. For signals at the chosen drift rate, there will be no sensitivity loss, and signals that are nearby in drift rate space can be searched for incoherently (with the same sensitivity loss considerations as discussed above), broadening the applicability of the technique. However, note that for lower loss tolerances at any given drift rate, the problem gets computationally heavier --- to minimize loss, it is necessary to dedisperse and save the entire array at every drift rate of interest.

This search, synchronized with planetary transits, is part of a class of search strategies aligned in time with particular astronomical events, as discussed in Section \ref{sec:intro}. For any such search strategy, we expect that the transmitter will likely not be transmitting continuously, but instead will only transmit at particular times, potentially with short durations. Therefore, searches for these kinds of signals benefit from observing with arrays that can do simultaneous off- and on-sources with multiple beams, instead of single-dish single-pixel instruments like the \ac{GBT} at L-band.

The short transmission times also make synchronization strategies more energy-efficient for the transmitter, similar to the argument made by \citet{gray2020intermittent}. Assume that the energy required to transmit for a full exoplanetary orbit is $E$. For the exoplanets in our sample, transmitting only during transit (as seen by a single receiver, in this case Earth), costs only ~1--10\% of $E$. If the transmitter instead only signalled during the 5 minutes around the mid-transit point (corresponding to the length of one of the observations in this campaign), the energy cost would be 0.1--0.01\% of $E$. Another strategy to improve energy efficiency could be to use a transmitter with a very large effective area, leading to an extremely narrow beam. In this strategy, the tiny beam size would cause a short flash right at mid-transit as seen from Earth. Having such a tiny angular beam size only works well if the observer knows exactly when to expect the signal, so it pairs well with transit synchronization. To an order-of-magnitude, using exoplanetary properties consistent with our 12 exoplanet sample, we can imagine a continuously-emitting transmitter with an effective area equivalent to an exoplanet's projected area. Before, we assumed the effective area would be the same as that of the \ac{GBT}, so the new effective area is now a billion times larger. This factor propagates, leading to such a small beamsize that a transmitted signal would be visible for a few milliseconds at mid-transit, and could be sent with a transmitter six orders-of-magnitude less powerful than Arecibo was. This paper does not account for an optimization this extreme, but it could be an interesting avenue for more specialized searches in the future.

\section{Conclusion}
\label{sec:conclusion}

In this work, we describe the first radio technosignature search that pre-planned observations to synchronize with exoplanets during their transits, in a survey of a dozen exoplanets in the Kepler field. Using 6 hours of L-band (1.1--1.9\,GHz) data taken with the \ac{BL} bankend on the \ac{GBT}, we performed a \ac{SETI} search using the narrowband signal search code \texttt{turboSETI}. We chose a maximum drift rate of $\pm$614.4\,Hz/s --- the first modern radio technosignature project to encompass such extreme drift rates --- in order to account for the full range of drifts that could be produced from known exoplanetary systems. We also chose a low \ac{S/N} of 5. With these parameters, the algorithms flagged $\sim 2.53 \times 10^6$ hits, which were then temporally filtered with an on-off method into $3.4 \times 10^5$ events. Many of these events could be attributed to \ac{GPS} satellite downlinks.

Thirteen citizen scientists volunteered their time to assist the science team with the further filtering of the \texttt{turboSETI} events. From this process, the list of signals-of-interest was reduced to a few thousand signals that appeared to be either transient or sky-localized. We further removed signals that appeared in multiple targets, or that were identical to signals proven to be \ac{RFI}, reducing the pool further to 479 signals-of-interest. Upon investigating these 479 signals in the context of the entire observing session, we determined that only two remained as signals-of-interest: one at 1749\,MHz in Kepler-1332b, and the other from 1040-1438\,MHz in Kepler-842b. 

These signals do not rise to the level of even ``candidate'' technosignature signals because there is no proof that they are spatially isolated (and they are consistent with anthropogenic \ac{RFI}), and so do not warrant follow-up with the rigor described in \citep{sheikh2021analysis}. Reobservation of these targets during transits, with a multibeam instrument such as the \ac{ATA}, will conclude the experiment and, if nothing is found, complete the null result we report here. We hope that the ``new ground'' that we have broken in radio technosignature parameter space will be extended by more synchronized \ac{SETI} searches in the future, across many more instruments and teams.

\begin{acronym}
\acro{ETA}{Extraterrestrial Agents}
\acro{GBT}{Green Bank Telescope}
\acro{BL}{Breakthrough Listen}
\acro{RFI}{Radio Frequency Interference}
\acro{S/N}{Signal-to-Noise Ratio}
\acro{SETI}{Search for Extraterrestrial Intelligence}
\acro{FCC}{Federal Communications Commission}
\acro{GPS}{Global Positioning System}
\acro{PSPSC}{Penn State Pulsar Search Collaboratory}
\acro{EIRP}{Equivalent Isotropic Radiated Power}
\acro{ATA}{Allen Telescope Array}
\end{acronym}

\begin{acknowledgements}

This work would not have been possible without the following citizen scientists: Killian Cook, Anish Doshi, Gus Eberlein, Rhett Gentile, Jordan Hanner, Shara Hussain, Matthew LaFountain, Yika Luo, Cole Penkunas, Livia Seifert, Nate Smith, Valeria Ventura, and James Wu. 

This material is based upon work supported by the Green Bank Observatory which is a major facility funded by the National Science Foundation operated by Associated Universities, Inc. We acknowledge Ron Maddalena for GBT observing and scientific assistance.

S.Z.S. acknowledges that this material is based upon work supported by the National Science Foundation MPS-Ascend Postdoctoral Research Fellowship under Grant No. 2138147NSF. CIC acknowledges support by NASA Headquarters under the NASA Earth and Space Science Fellowship Program through grant 80NSSC18K1114 and the Pennsylvania State University's Bunton-Waller program. C.G. acknowledges the support of the Pennsylvania State University, the Penn State Eberly College of Science and Department of Astronomy \& Astrophysics, the Center for Exoplanets and Habitable Worlds and the Center for Astrostatistics.

This paper is a result of the class project for the 2020 graduate course in SETI at Penn State. We acknowledge Alan Reyes for participation in the class project. The Pennsylvania State University campuses are located on the original homelands of the Erie, Haudenosaunee (Seneca, Cayuga, Onondaga, Oneida, Mohawk, and Tuscarora), Lenape (Delaware Nation, Delaware Tribe, Stockbridge-Munsee), Shawnee (Absentee, Eastern, and Oklahoma), Susquehannock, and Wahzhazhe (Osage) Nations. As a land grant institution, we acknowledge and honor the traditional caretakers of these lands and strive to understand and model their responsible stewardship. We also acknowledge the longer history of these lands and our place in that history. 

Breakthrough Listen is managed by the Breakthrough Initiatives, sponsored by the Breakthrough Prize Foundation. The Center for Exoplanets and Habitable Worlds and the Penn State Extraterrestrial Intelligence Center are supported by the Penn State and the Eberly College of Science. This research made use of Astropy,\footnote{http://www.astropy.org} a community-developed core Python package for Astronomy \citep{astropy:2013, astropy:2018}.  This research has made use of NASA's Astrophysics Data System Bibliographic Services.
\end{acknowledgements}

\facilities{GBT}

\software{
Astropy \citep{astropy:2013, astropy:2018}, 
Numpy \citep{Harris:2020},
Matplotlib \citep{Hunter:2007},
\texttt{blimpy} \citep{breakthrough2019blimpy, Price2019b},
\texttt{turboSETI} \citep{enriquez2019turboseti}.
}

\bibliography{references_SZS, references_SK}

\appendix
\section{Observation Log}

\begin{table}[ht]
    \centering
    \caption{The observation log from this project's Green Bank Telescope observations. The scan ID number is shown in the first column, the target is shown in the second column, the transit synchronization is shown in the third column, and any additional notes are shown in the fourth column. Scan IDs are mostly consecutive, but missing ID numbers indicate scans which did not take data. Targets prefixed with a ``K'' indicate a ``Kepler'' system, with the letter suffix indicating which planet in the system was predicted to be transiting during our observing window. The relationship of the observation time to the transit time is indicated by letters: I-M indicates an observation which occurred between a transit's ingress and midpoint, M indicates an observation which contained the transit midpoint, and M-E indicates an observation which occurred between a transit's midpoint and egress (blank entries occurred outside of transit). Scan 0051 is missing a node of data (about 1/8th of the bandwidth) due to faulty data collection, encompassing frequencies from 1.25--1.30 GHz (entirely within the notch filter).}
    \begin{tabular}{c c c c||c c c c }
    \hline
    \hline
    Scan ID & Target & Relation to Transit & Notes & Scan ID & Target & Relation to Transit & Notes\\
    \hline
    0006 & 3C295            & non-planet obs.       & Quasar    & 0033 & K723b            & I-M       & \\
    0007 & 3C295            & non-planet obs.      & Quasar    & 0034 & K1164b           & M-E       & \\
    0009 & Boyajian's Star  & non-planet obs.      & Off-source   &    0035 & K723b            & M-E       & \\
    0010 & K992b            & M         & &     0036 & K537b            & M         & \\
    0011 & K738b            & I-M       & &    0037 & K723b            & I-M       & \\
    0012 & K992b            & M-E       & &    0038 & K537b            & M-E       & \\
    0013 & K738b            & I-M       & &    0039 & K723b            & I-M       & \\
    0014 & K1039b           & M         & &    0040 & K537b            & M-E       & \\
    0015 & K1039b           & M-E       & &    0041 & K723b            & I-M       & \\
    0016 & K732c            & M         & &    0042 & K723b            & M         & \\
    0017 & K1039b           & M-E       & &    0043 & K1332b           &           & \\
    0018 & K738b            & M         & &    0044 & K723b            & M-E       & \\
    0019 & K738b            & M-E       & &    0045 & K1332b           &           & \\
    0020 & K732c            & M-E       & &    0046 & K723b            & M-E       & \\
    0021 & K1164b           & I-M       & &    0048 & K723b            & M-E       & \\
    0022 & K732c            &           & &    0049 & K446b            & I-M       & \\
    0023 & K1164b           & I-M       & &    0050 & K723b            & M-E       & \\
    0024 & K732c            &           & &    0051 & K446b            & I-M       & missing node \\
    0025 & K1164b           & I-M       & &    0052 & K446b            & M         & \\
    0026 & K1053b           & M         & &    0053 & K1222b           & I-M       & \\
    0027 & K738b            & M-E       & &    0054 & K1222b           & M         & \\
    0028 & K1164b           & M         & &    0055 & K842b            & I-M       & \\
    0029 & K1053b           & M-E       & &    0056 & K842b            & M         & \\
    0030 & K738b            & M-E       & &    0057 & K723b            & M-E       & \\
    0031 & K1332b           & M         & &    0058 & K723b            & M-E       & \\
    0032 & K1164b           & M-E       & &    0059 & B0329+54         & non-planet obs.       & Pulsar\\
    \hline
    \end{tabular}

    \label{tab:observation_table}
\end{table}

\section{Mid-resolution plots for the two signals-of-interest}

\begin{figure}
    \centering
    \includegraphics[width=\textwidth]{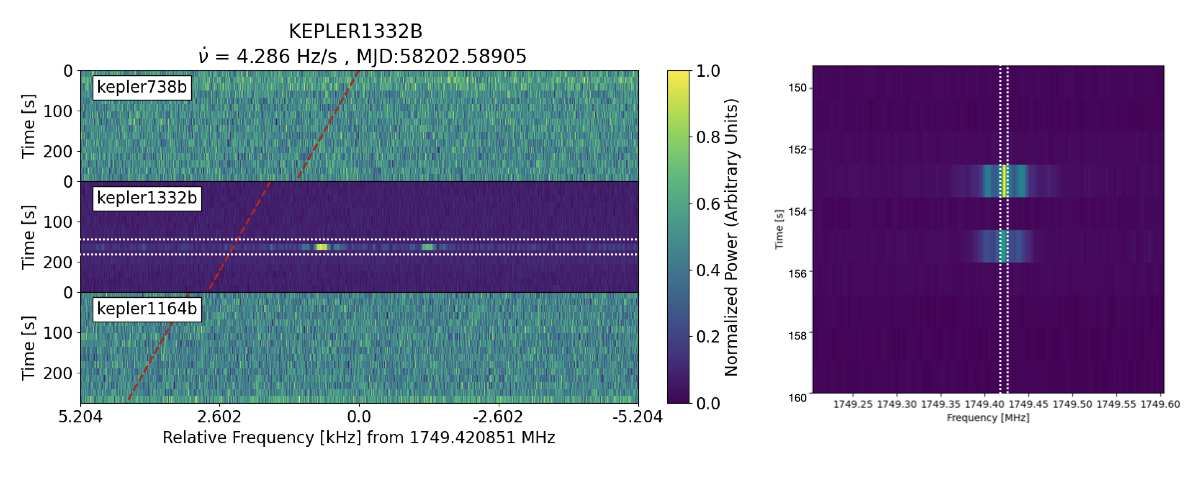}
    \caption{The signal-of-interest in Kepler-1332b data shown in the high frequency-resolution data (left) and the mid-resolution data (right). The regions of data shown are related as indicated by the white-dashed rectangles, with the right plot zoomed-out in frequency and zoomed-in in time. The signal-of-interest is actually composed of two short ($<$1 second) pulses.}
\end{figure}

\begin{figure}
    \centering
    \includegraphics[width=\textwidth]{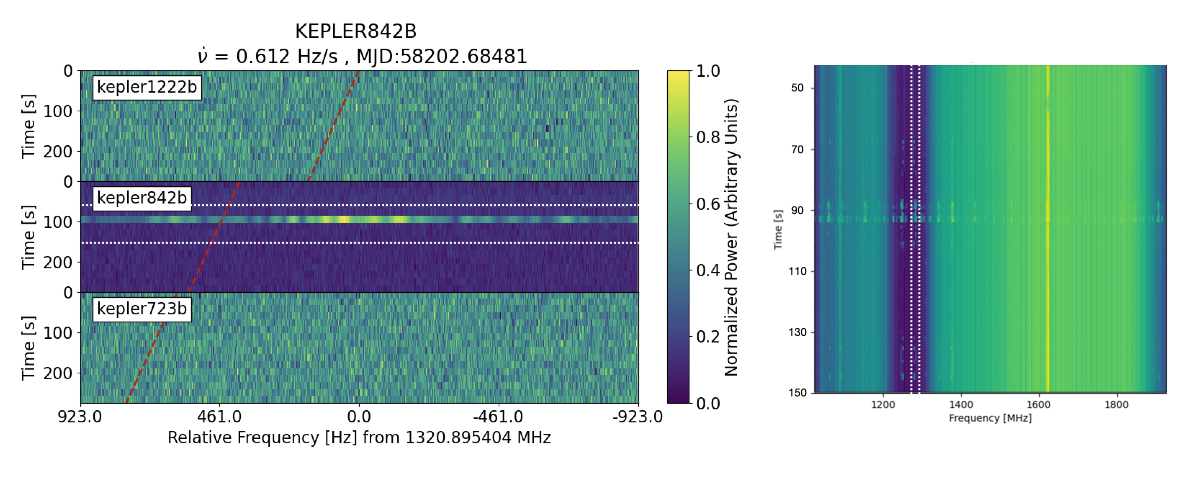}
    \caption{The signal-of-interest in Kepler-842b data shown in the high frequency-resolution data (left) and the mid-resolution data (right). The regions of data shown are related as indicated by the white-dashed rectangles, with the right plot zoomed-out in frequency and zoomed-in in time. The signal-of-interest has a distinctive, periodic structure in frequency and is present across the bandwidth. This behaviour is characteristic of \ac{RFI}.}
\end{figure}

\listofchanges

\end{document}